# Application of the New Form of the Semiclassical Quantization Condition

N.N.Trunov[1]

*D.I.Mendeleyev Institute for Metrology*

*Russia, St. Petersburg. 190005 Moskovsky pr. 19*

(Dated: December 19, 2008)

**Abstract:** We demonstrate how a certain new form of the quantization condition proposed earlier can be used outside the class of potentials for which this form ensures exact spectra. Taking this form as a base we get an improved interpolating condition for the potential wells with coinciding at $\infty$ and $-\infty$ asymptotical values. Accuracy of the above conditions is discussed.

## 1. Properties of the quantization condition

Without loss of generality we can write the equation determining energy levels $\varepsilon_n$ in a potential well $V(x)$ in the following form:

$$\Phi(\varepsilon) = \frac{1}{\pi\beta}\int \sqrt{\varepsilon - V}\, dx = n + \tfrac{1}{2} + \delta \qquad (1)$$

with the unknown yet function $\delta$ which ensures the exact spectrum. Here $n = 0, 1, 2...$, the integral is taken between turning points and

$$\beta^2 = \hbar^2/2m. \qquad (2)$$

---

[1] Electronic address: trunov@vniim.ru

The common popular Semiclassical condition corresponds to $\delta \equiv 0$. It is known that such condition is exact only for few potentials. A power series expansion

$$\delta = \sum_{k=1}^{\infty} \delta_k$$
$$\delta_k = f_k(\varepsilon)\beta^{2k-1}$$
(3)

was formally built for $\delta$ but is used very seldom since $\delta_k$ are very cumbersome for $k > 1$, $\delta_1$ see in (5).

In the previous paper [1] we have proposed the following expression for $\delta$:

$$\delta = \frac{2\delta_1}{1+\sqrt{1+16\delta_1^2}};$$
(4)

$$\delta_1 = \frac{\beta}{24\pi}\frac{\partial^2}{\partial\varepsilon^2}\int \frac{dx}{\sqrt{\varepsilon-V}}\left(\frac{dV}{dx}\right)^2.$$
(5)

The quantization condition (1) with $\delta$ (4) is exact for all potentials which may be expressed by means of an auxiliary function $s(x)$ as

$$V(x) = A^2 s^2 + Bs + C$$
$$\sigma \equiv \frac{ds}{dx} = a_2 s^2 + a_1 s + a_0$$
(6)

It should be stressed that (5) is not a given ad hoc class but embraces all the potentials used in handbooks as model and reference ones for the semiclassical approximation as well as for exact solutions [1,2]. (Excepting those potentials for which spectra are roots of some transcendental functions).

This fact means that real interesting potentials are situated in some vicinity (measured in a suitable metric) of potentials (6), for which

$$\delta_1 = \frac{\beta a_2}{8A}$$
(7)

as well as $\delta$ does not depend on $\varepsilon$. Not that (7) is invariant under the transformation $V \to V - C$, $s \to s + const$, in many cases it is convenient to choose $V = A^2 s^2$ (with the same $a_2, A$).

Let's show the limiting values:
$$\delta = \delta_1, \qquad |\delta_1| \to 0 \qquad (8)$$

$$\delta = \frac{\operatorname{sgn}\delta_1}{2} + \frac{1}{8\delta_1}, \qquad |\delta_1| \to \infty. \qquad (9)$$

Exactness of the quantization condition (1), (4) for all the potentials (6) means that in some vicinity of these potentials though (4), (5) is already not exact, it rests a good approximation for $\delta$ and may serve as a base for a closer definition of $\delta$.

As a dimensionless parameter indicating the deviation of a studied potential from the class (6) may serve

$$\gamma = \frac{d\delta}{dn} \qquad (10)$$

as well as other parameters, which turn into zero simultaneously with (10), i.e. for (6).

One may also say that (4), (5) is the exact sum of (3) for all potentials (6) and outside (6) is an approximate sum as a result of the partial summarization of all terms $\delta_k$ with all values of $k$, i.e. with all powers $\beta^{2k-1}$. Corresponding error is small if $|\gamma| \ll 1$.

## 2. An improved approximation of $\delta$

In the present section we study how to precise (4), (5) for the potential wells $V(x)$ with the coinciding asymptotics

$$V(x) \to U, \quad |x| \to \infty, \qquad (11)$$

$$V(0) = 0.$$

A known example is

$$V = A^2 s^2, \quad s = th(x), \quad A^2 = U. \qquad (12)$$

It is easy to calculate for (12)

$$\sigma = \frac{ds}{dx} = 1 - s^2 \qquad (13)$$

so that (12) belongs to the class (6) and (4), (5) are exact, $\gamma = 0$.

It is known that for each potential (12) – not only for (6) – the lower bound state with $n = 0$ exists for an arbitrary value of $U$, even if $A^2 = U \to 0$. In this case $\varepsilon_0 = U$ to the approximation of $U^2 \ll U$.

The value of $\delta_1$ (5) tends to infinity when $A$ tends to zero. This property remains in force for any potential with asymptotics (11), at least in a vicinity of (6).

Now we intend to get an improved approximate form of $\delta$ instead of (4) using two limiting cases: small $\delta_1$ when $\delta = \delta_1$ for any $V(x)$ and very large $|\delta_1|$ which is reached for $U \to 0$. It'll be convenient to use very effective two-point Pade approximations, but we have to keep radicals in the denominator of (5). Thus we propose the following improved expression for $\delta$ with one parameter $q$:

$$\delta = \frac{2\delta_1}{q + \sqrt{(2-q)^2 + 16\delta_1^2}} ; \qquad (14)$$

$$q = -\frac{8\Phi(U)}{\delta_1}. \qquad (15)$$

Really we have in (14) $\delta = \delta_1$ at any $q$ and small $\delta_1$ but at very large $|\delta_1|$, $\delta_1 < 0$

$$\delta = -\frac{1}{2} + \Phi(U). \qquad (16)$$

Now, substituting into (1) for the lower state $n = 0$, $\varepsilon = U$ at very small $U$ and $\delta$ (16) we make sure that left and right sides of (1) actually coincide.

It is easy to calculate that for all potentials (11), (6) $q = 1$ independently on values $U$ (and $\varepsilon$). Thus the vicinity of (6) for potentials (11) may be determined as $|\mu| \ll 1$,

$$\mu = 1 - q. \qquad (17)$$

Note that we can also use $q$ in a simplified manner: since $\Phi(U)$ is easier to calculate than $\delta_1$ it is possible approximately put $q=1$ and then replace $\delta_1$ in (5) with $\Phi$ by means of (15) with $q=1$.

## 3. Accuracy of the semiclassical quantization

An additional shift $\delta$ in (1) gives a natural measure of the accuracy of (1). As it follows from the above, the simple condition with $\delta = 0$ is exact only for the subset of potentials (6) with $a_2 = 0$, see (7). All usually mentioned in handbooks qualitative explanations of exact semiclassical spectra (with $\delta = 0$) not appealing to (6) are incorrect.

Taking into account some approximate $\delta$ instead of the exact $\delta_{ex}$ we obtain an error $\delta - \delta_{ex}$ proportional to $\gamma$ (10) or $\mu$ (17), which are small for most of interesting potentials.